\documentclass[epj]{webofc}
\usepackage[varg]{txfonts}   
\usepackage{hyperref}
\newcommand{\ptn}{p_{\rm{T}}}

\newcommand{\pp}{pp}
\newcommand{\pPb}{p--Pb}

\newcommand{\PbPb}{Pb--Pb}
\newcommand{\dNdeta}{\langle dN_{ch}/d\eta\rangle}
\woctitle{XLVI International Symposium on Multiparticle Dynamics}
\begin{document}
\title{Correlations, multiplicity distributions, and the ridge in pp and p-Pb collisions}

\author{Alice Ohlson\inst{1}\fnsep\thanks{\email{alice.ohlson@cern.ch}} for the ALICE Collaboration}

\institute{Ruprecht-Karls-Universit\"{a}t Heidelberg, Heidelberg, Germany}

\abstract{%
Measurements made by the ALICE Collaboration of single- and two-particle distributions in high-energy \pp{} and \pPb{} collisions are used to characterize the interactions in small collision systems, tune models of particle production in QCD, and serve as a baseline for heavy-ion observables.  The measurements of charged-particle multiplicity density, $\dNdeta$, and multiplicity distributions are shown in \pp{} and \pPb{} collisions, including data from the top center-of-mass energy achieved at the Large Hadron Collider (LHC), $\sqrt{s}$ = 13 TeV.   Two-particle angular correlations in \pPb{} collisions are studied in detail to investigate long-range correlations in pseudorapidity which are reminiscent of structures previously thought unique to heavy-ion collisions.  
}
\maketitle
\section{Introduction}
In high-energy hadronic collisions, studies of inclusive single-particle distributions are used to investigate particle production in QCD.  The charged-particle multiplicity density in \pp{} and \pPb{} collisions is measured by ALICE over a range of centre-of-mass energies, including the top LHC energy of $\sqrt{s} = 13$ TeV.  The multiplicity distributions are also shown, and all the experimental data is compared with Monte Carlo models.  Since the produced multiplicity is dominated by soft (low-momentum) particle production, which is in the non-perturbative regime of QCD, these measurements can be used to further constrain and tune models.  

Beyond single-particle inclusive measurements, two-particle correlation studies have yielded surprising results in small collision systems, showing the presence of correlations between particles over large ranges in pseudorapidity in high-multiplicity \pp{} and \pPb{} collisions.  These correlations are reminiscent of features observed in heavy-ion collisions where they are commonly attributed to anisotropic flow ($v_n$).  The transverse momentum ($\ptn$), pseudorapidity ($\eta$), and particle species dependence of $v_2$ in \pPb{} collisions has been measured in ALICE.  In particular, in the analysis of correlations between forward muons and mid-rapidity charged hadrons it is possible to measure the $v_2$ for large values of pseudorapidity in both the proton-going and Pb-going directions.  These observations will be used to deepen our understanding of possible collective effects in small collision systems and their implications for heavy-ion physics.  

\section{ALICE detector}

The main subsystems of the ALICE detector~\cite{ALICEdet} used in the analyses reported here are: the Inner Tracking System (ITS), the Time Projection Chamber (TPC), the Forward Muon Spectrometer (FMS), and the V0 system.  The ITS, used for tracking and vertex reconstruction, consists of six layers of silicon detectors; the innermost two layers comprise the Silicon Pixel Detector (SPD).  Short track segments (``tracklets'') can be reconstructed using only the SPD, and are used in the multiplicity density and muon-hadron correlations analyses below.  Information from the ITS and TPC can also be combined to fully reconstruct charged particle tracks.  Muons are detected in the FMS, which has a pseudorapidity coverage of $-4 < \eta < -2.5$.  The composition of parent particles of the detected muons depends on transverse momentum: at low $\ptn$ the muons predominantly come from weak decays of pions and kaons, while at high $\ptn$ the muons are largely the result of heavy flavor decays.  The V0 detectors, located at forward rapidity (the V0A at $2.8 < \eta < 5.1$ and the V0C at $-3.7 < \eta < -1.7$), are used for triggering and also to classify the overall event activity.  Symmetric pseudorapidity coverage can also be achieved by utilizing only two of the four rings in each V0 detector, the innermost two rings of the V0C ($-3.7 < \eta < -2.7$) and the outermost two rings of the V0A ($2.8 < \eta < 3.9$), as was done in the muon-hadron analysis below.  

\section{Multiplicity density}
\begin{figure*}[b!]
\centering
\includegraphics[width=0.40\linewidth]{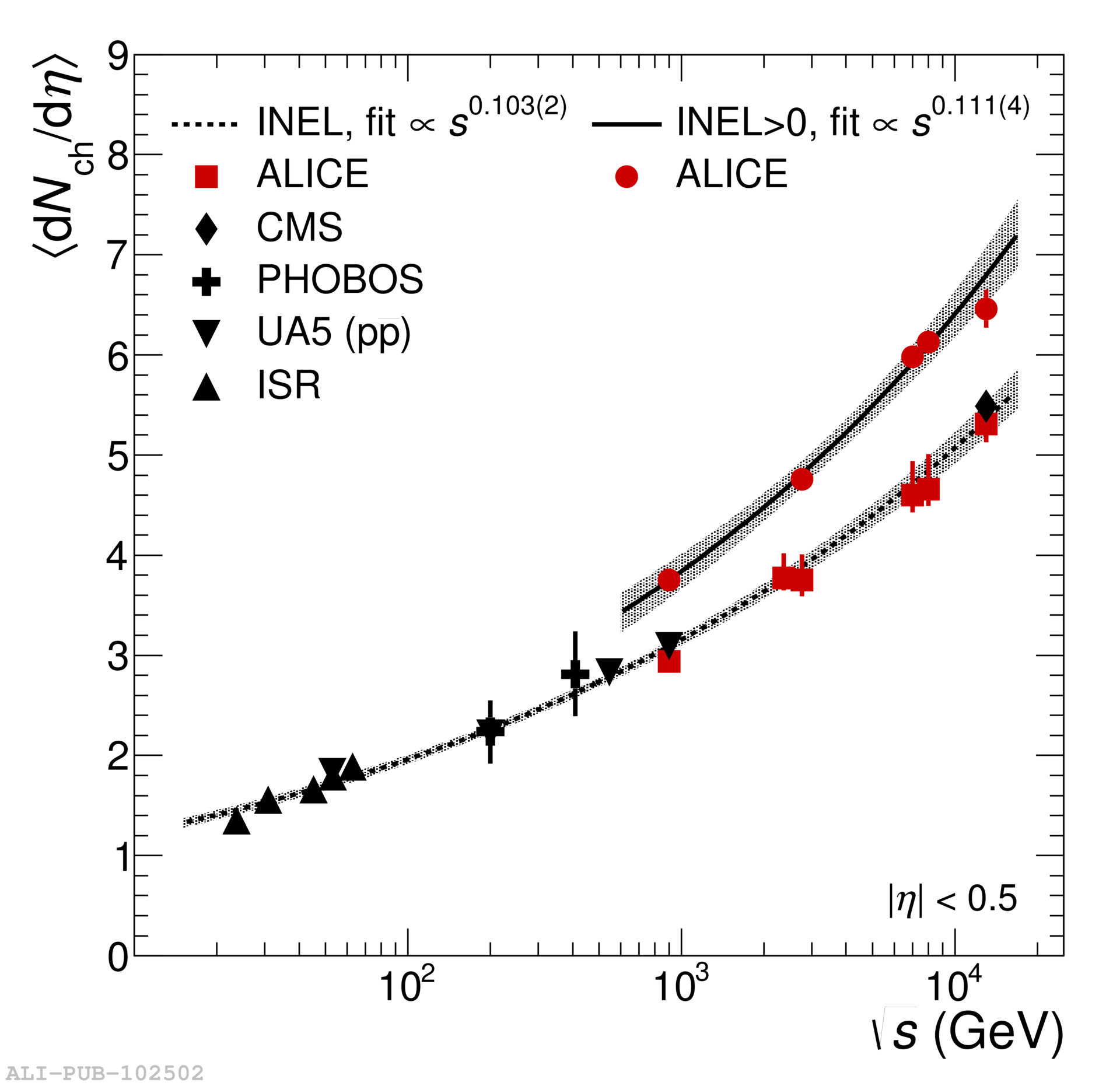}
\includegraphics[width=0.39\linewidth]{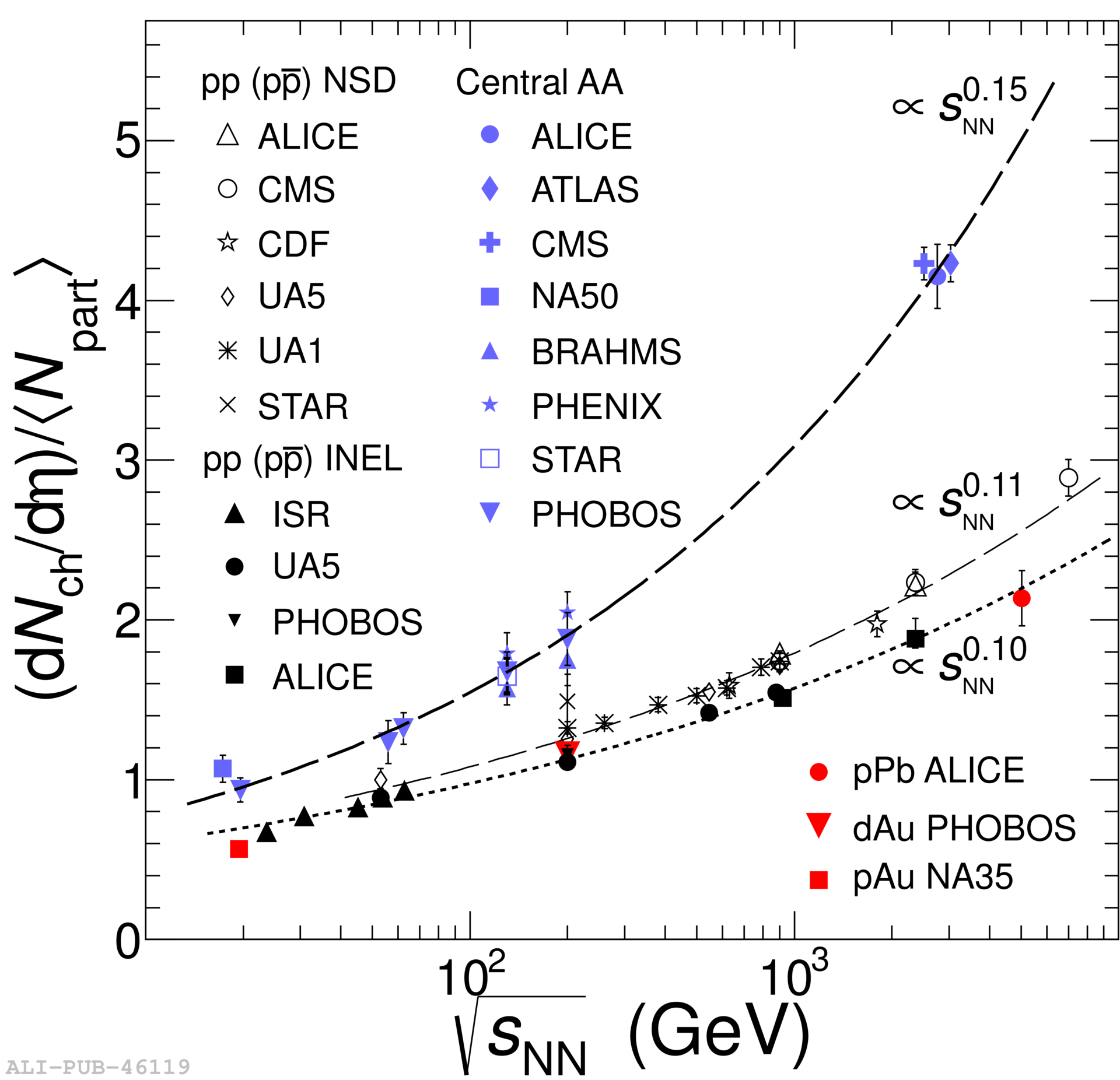}
\caption{The mid-rapidity charged-particle multiplicity density, $\dNdeta$, is shown as a function of center-of-mass energy for (left) \pp{} and (right) \pp{}, \pPb{}, and \PbPb{} collisions at the LHC~\cite{mult2016,multpPb}.}
\label{fig:fig1}
\end{figure*}

In ALICE, the charged-particle multiplicity density, $\dNdeta$ has been measured across a wide range in center-of-mass energy, at $\sqrt{s} =$ 0.9, 2.36, 2.76, 7, 8, and 13 TeV.  The multiplicity is measured in different classes of events, including inelastic events (`INEL'), inelastic events with at least one charged particle produced within $|\eta|<1$ (`INEL\textgreater 0'), and non-single-diffractive events (`NSD').  Figure~\ref{fig:fig1} shows the results for the INEL and INEL\textgreater 0 classes, which demonstrate power-law scaling with $\sqrt{s}$.  Results from \pPb{} collisions are also shown in Fig.~\ref{fig:fig1}~\cite{multpPb}.

\begin{figure*}[t!]
\centering
\includegraphics[width=0.715\linewidth]{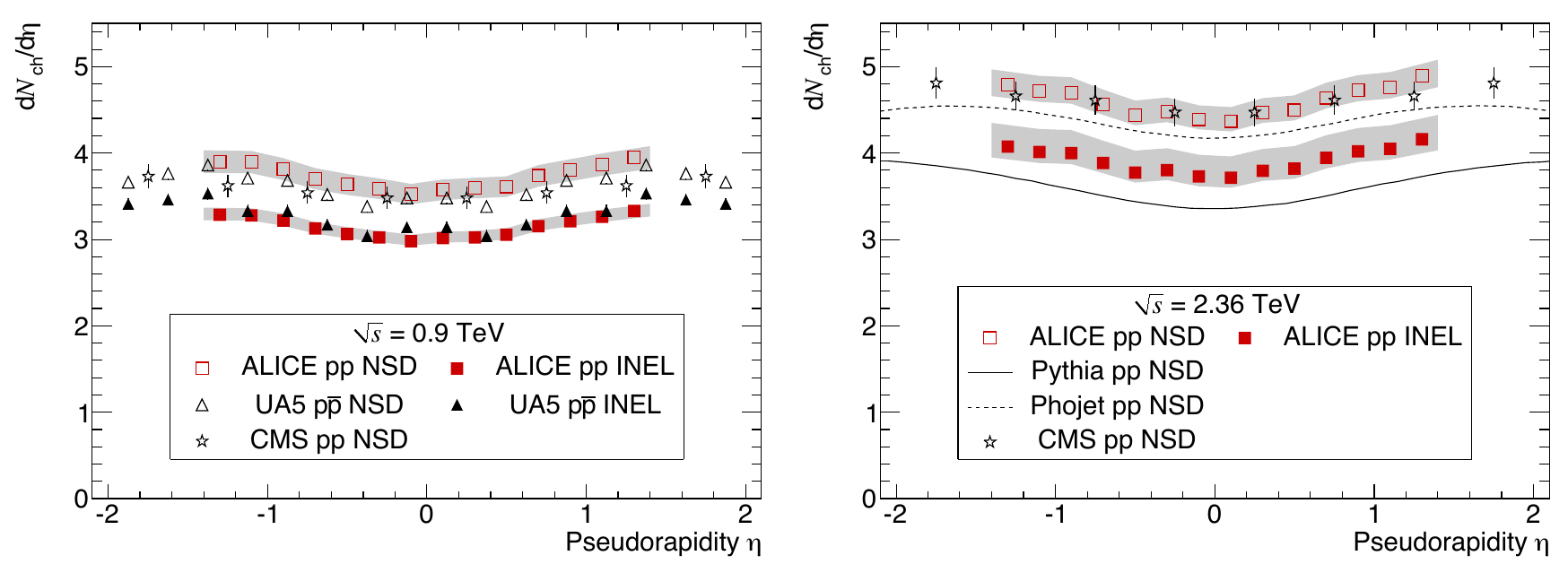}
\includegraphics[width=0.27\linewidth]{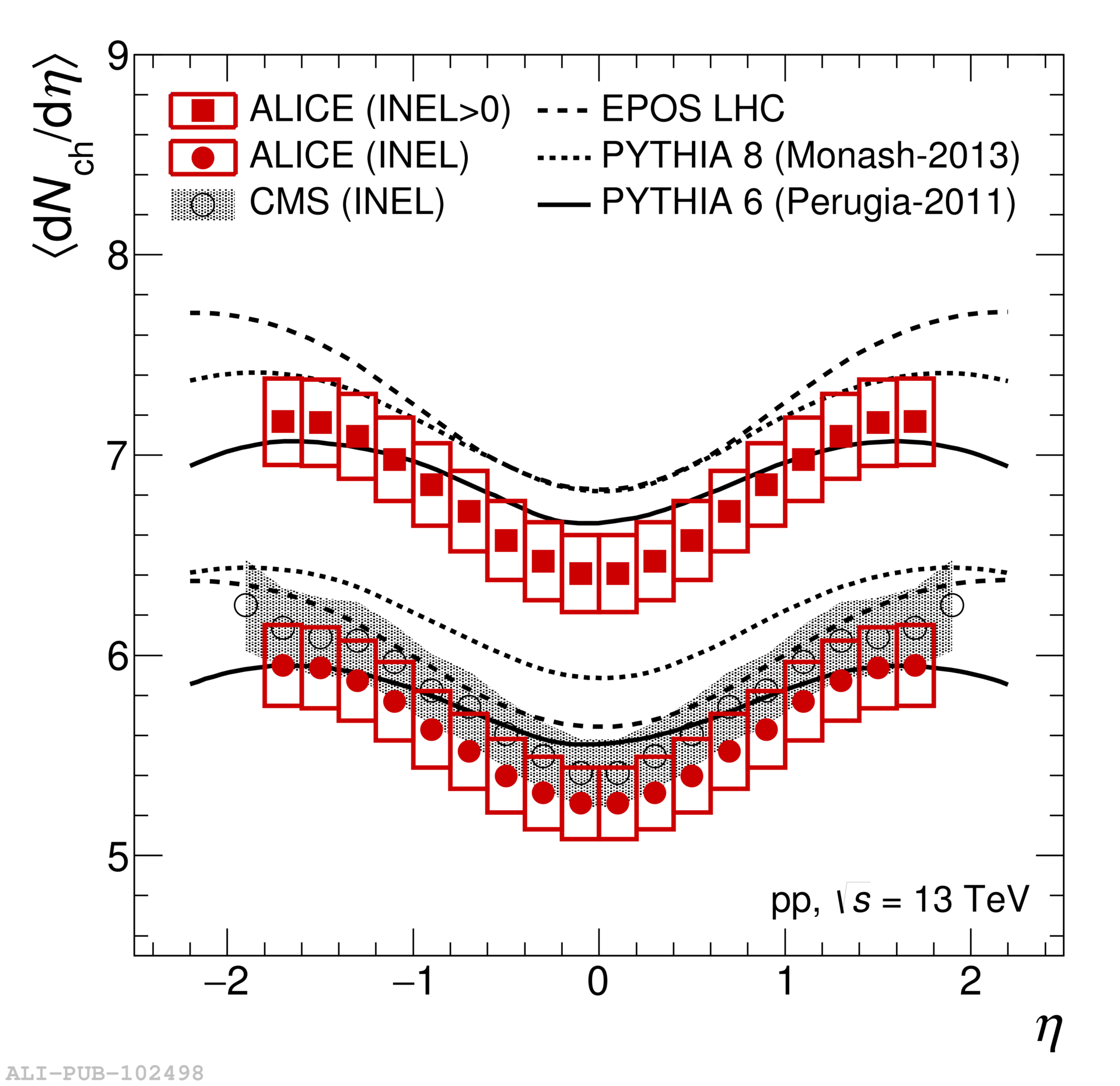}
\caption{$\dNdeta$ vs $\eta$ in \pp{} collisions is shown at $\sqrt{s} = 0.9$ (left), 2.36 (center), and 13 TeV (right)~\cite{mult2010,mult2016}.}
\label{fig:fig2}
\end{figure*}

Additionally, the charged particle multiplicity density has been measured as a function of pseudorapidity in INEL and NSD events at $\sqrt{s}$ = 0.9 and 2.36 TeV, and in INEL and INEL\textgreater 0 events at 13 TeV, as shown in Fig.~\ref{fig:fig2}.  The distributions are also compared to multiple Monte Carlo (MC) models including PYTHIA 6~\cite{pythia6}, PYTHIA 8~\cite{pythia8}, PHOJET~\cite{phojet}, and EPOS LHC~\cite{eposlhc}.  It can be observed in Fig.~\ref{fig:fig2} that the model in best agreement with the $\sqrt{s}$ = 13 TeV data is PYTHIA 6.  These results will be used for further tuning of the Monte Carlo generators.  

\section{Multiplicity distributions}

\begin{figure*}[t!]
\centering
\includegraphics[width=0.66\linewidth]{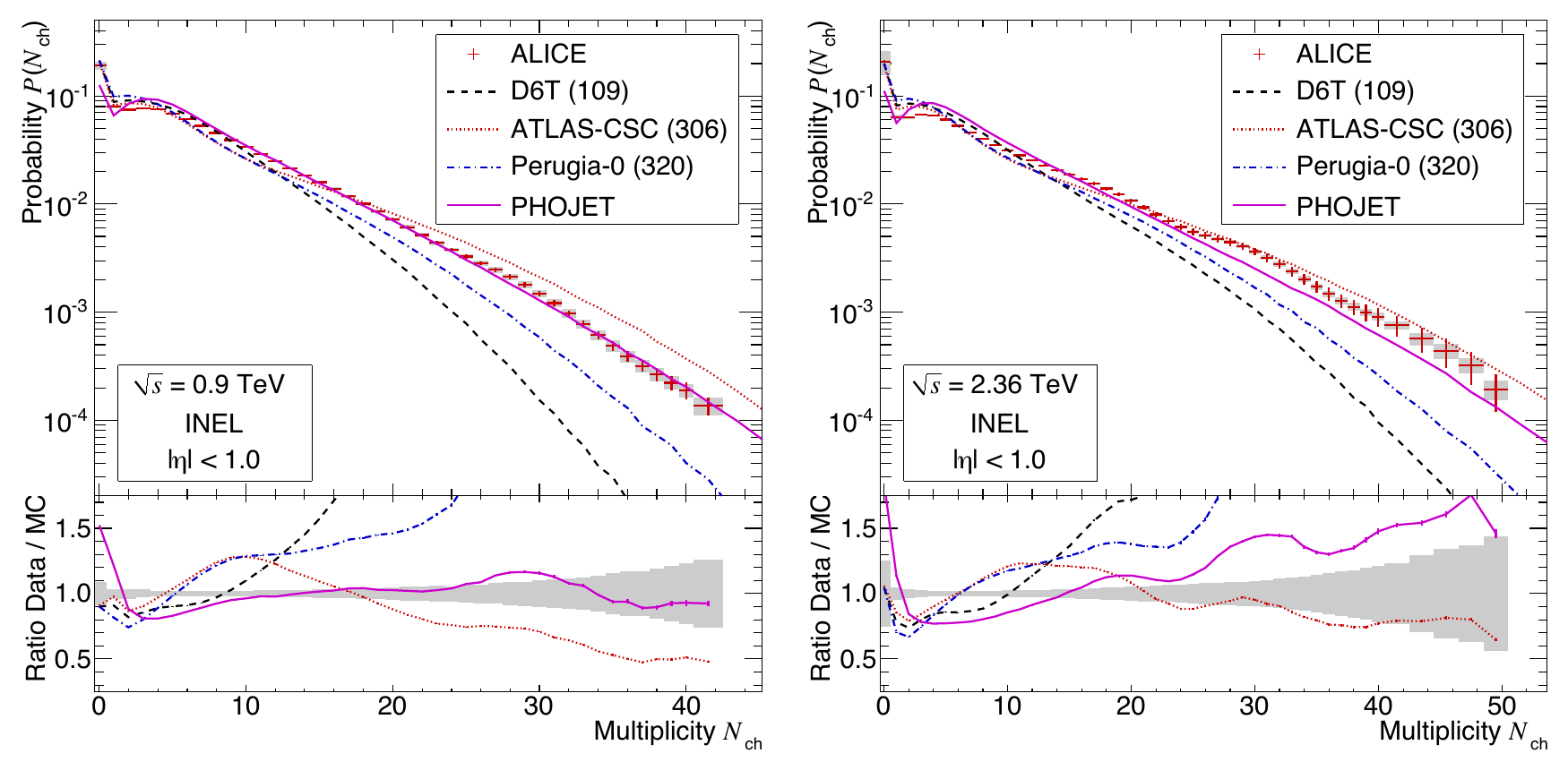}
\includegraphics[width=0.32\linewidth]{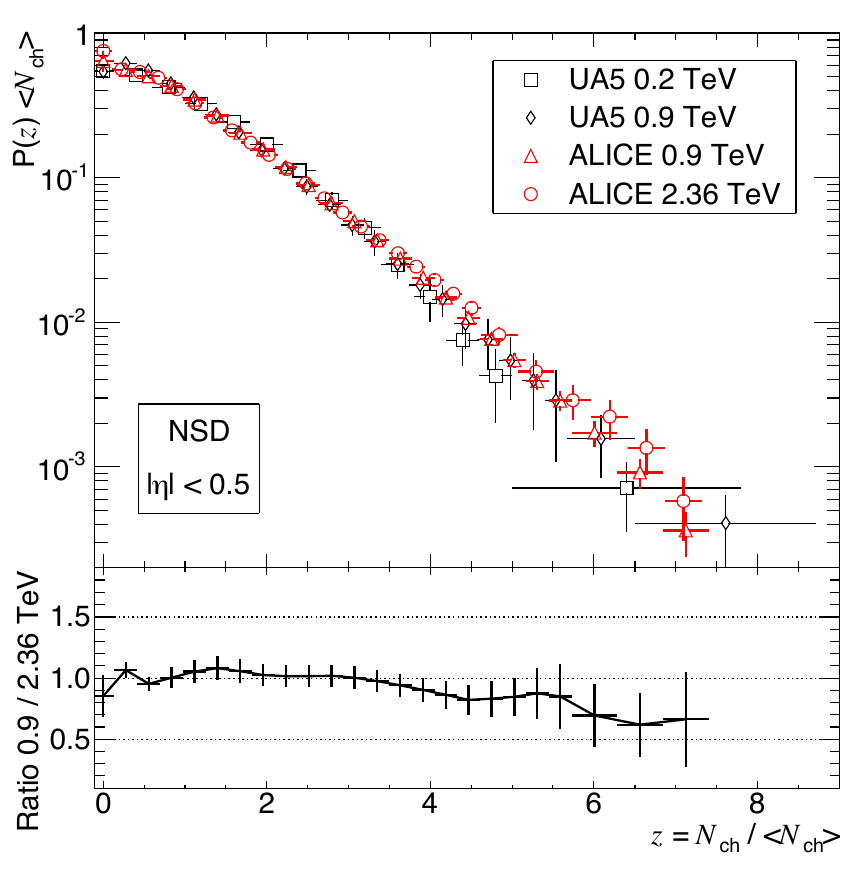}
\caption{The multiplicity distributions are shown at $\sqrt{s} = 0.9$ (left) and 2.36 TeV (center) with MC model comparisons.   KNO scaling is also shown (right).~\cite{mult2010}}
\label{fig:fig3}
\end{figure*}

The charged-particle multiplicity distributions, $P(N_{ch})$, were measured at $\sqrt{s}$ = 0.9 and 2.36 TeV, as shown in Fig.~\ref{fig:fig3}.  The experimental data were compared to results from PHOJET and three PYTHIA 6 tunes (Perugia-0, ATLAS-CSC, and D6T).  The best agreement with the data is achieved by PHOJET at $\sqrt{s}$ = 0.9 TeV and the ATLAS-CSC tune of PYTHIA 6 at $\sqrt{s}$ = 2.36 TeV.  Furthermore, the multiplicity distributions were scaled by the mean multiplicity to obtain the distribution of $z = N_{ch}/\langle N_{ch}\rangle$, also shown in Fig.~\ref{fig:fig3}.  The hypothesis that the distributions of $\langle N_{ch}\rangle P(z)$ are independent of center-of-mass energy is known as KNO scaling~\cite{kno}, and these experimental results indicate that KNO scaling holds up to approximately $z = 4$.  

\section{Two-particle correlations}

Two-particle angular correlations, which are distributions in relative azimuthal angle ($\Delta\varphi = \varphi_{trig}-\varphi_{assoc}$) and relative pseudorapidity ($\Delta\eta = \eta_{trig}-\eta_{assoc}$) between trigger and associated particles, are used to study many aspects of the physics of heavy-ion collisions, in particular jet fragmentation and collective effects.  In elementary collisions and small collision systems such as \pp{} they show characteristic features attributed to jet production, while in heavy-ion collisions the same jet features are observed in addition to structures around $\Delta\varphi = 0$ (nearside) and $\Delta\varphi = \pi$ (awayside) extended in $\Delta\eta$.  These long-range correlations, known as `ridges,' are often attributed to hydrodynamic flow behavior in the quark-gluon plasma (QGP) and are typically quantified by the coefficients of a Fourier cosine series, $v_n$.  

\begin{figure}[b!]
\centering \includegraphics[width=0.3\textwidth]{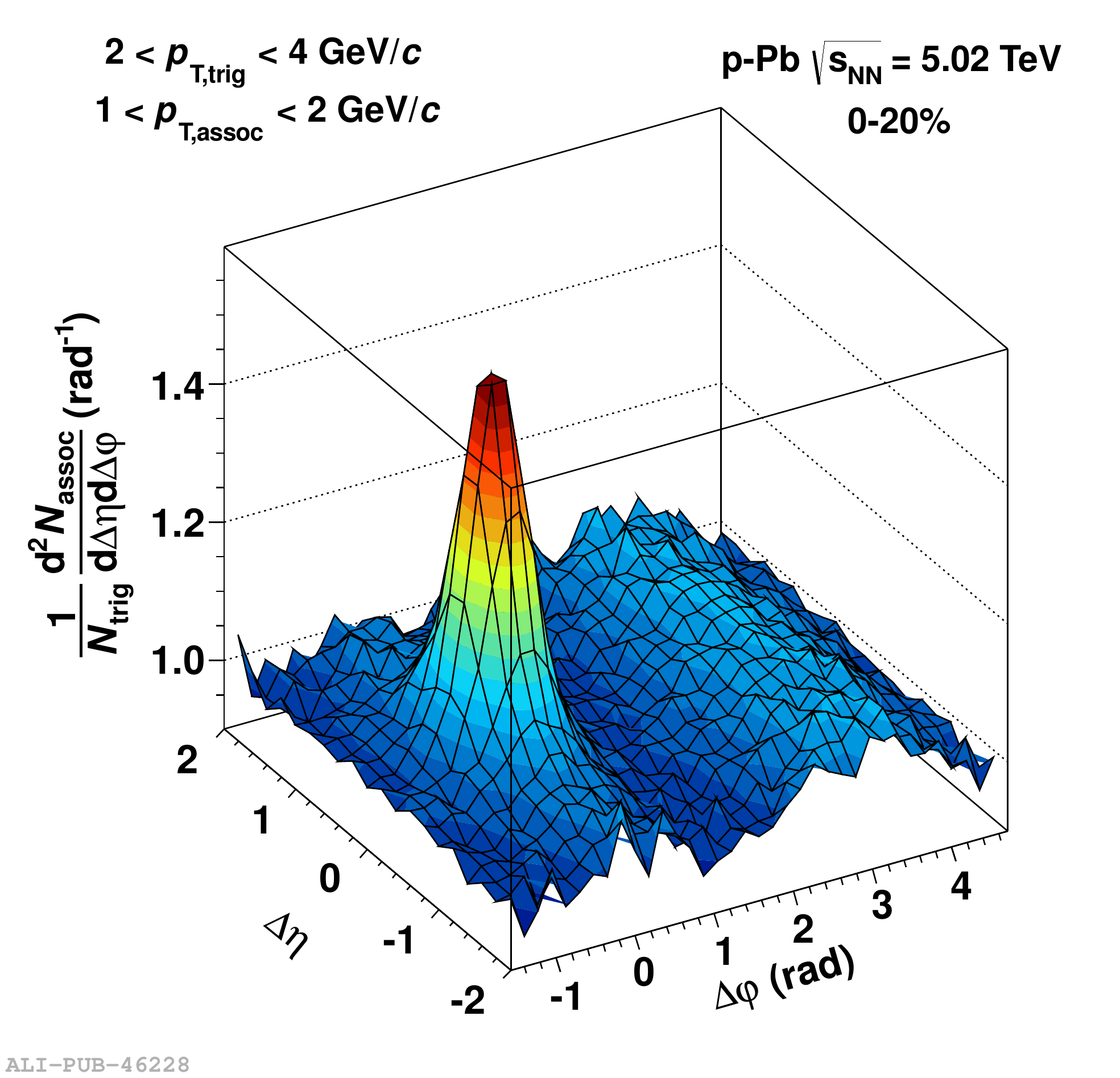}~
\centering \includegraphics[width=0.3\textwidth]{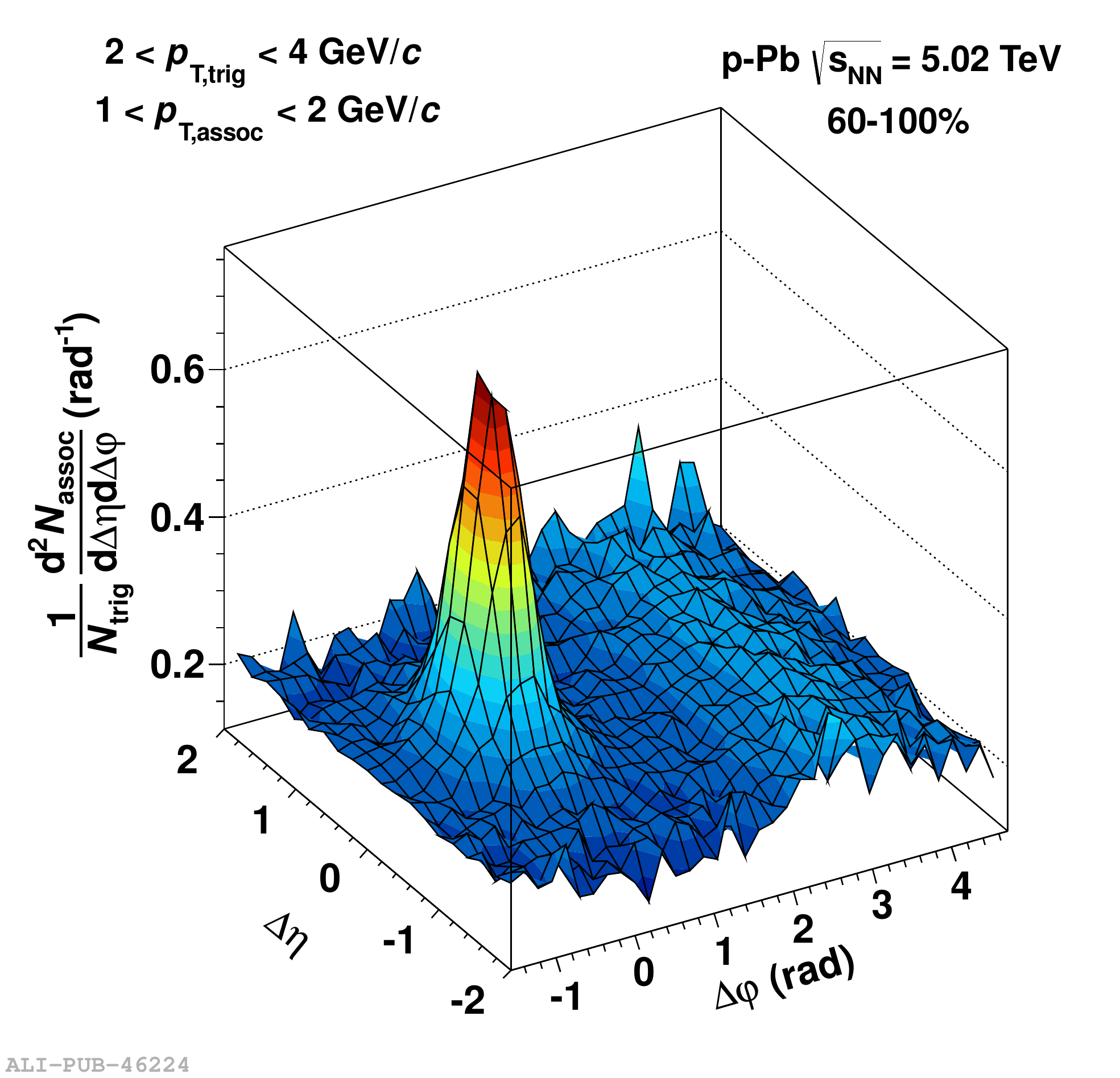}~
\centering \includegraphics[width=0.3\textwidth]{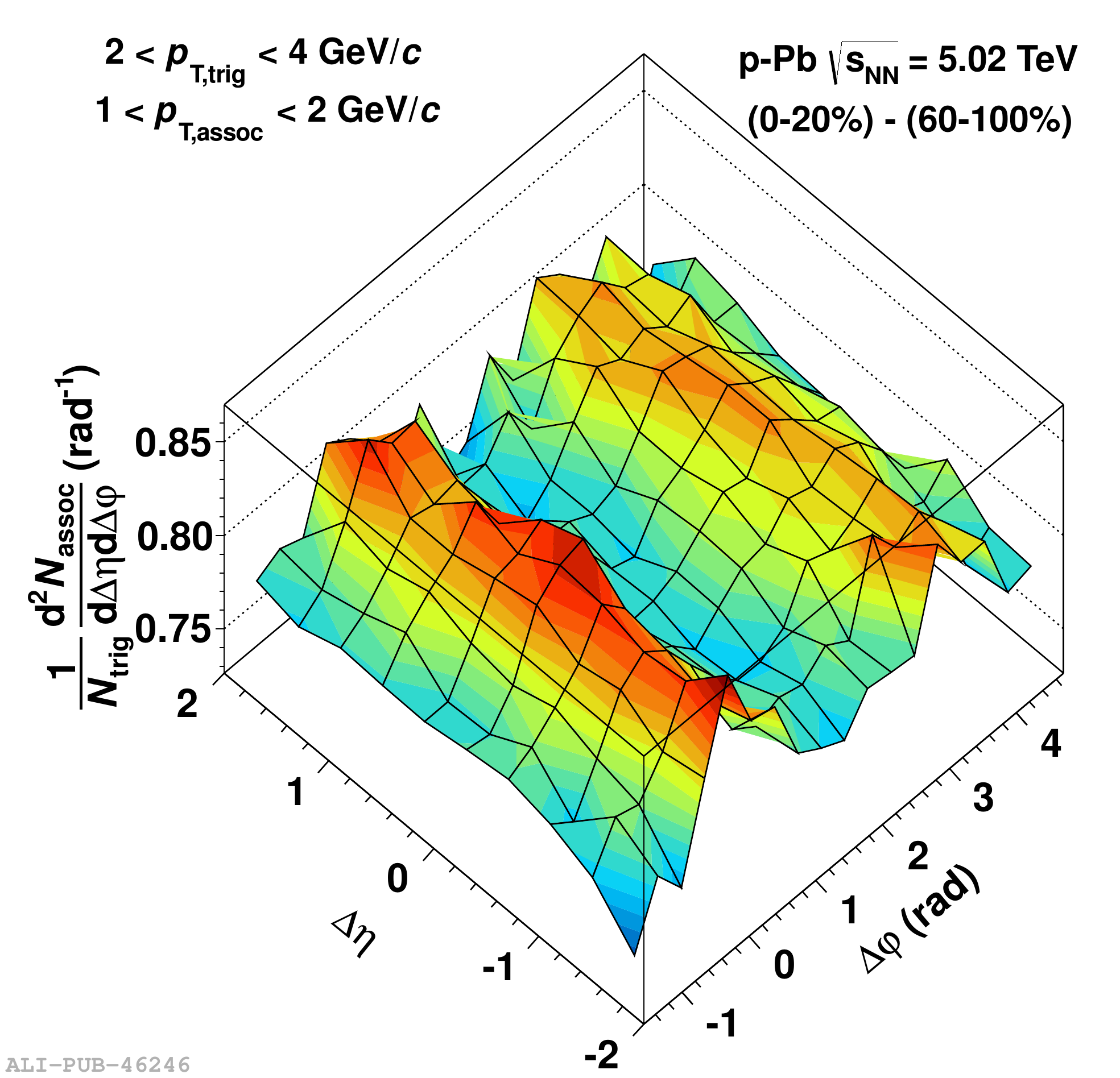}
\caption{The two-particle correlation functions in \pPb{} collisions show a nearside ridge in high-multiplicity collisions (left) while no ridge is visible in low-multiplicity collisions (center).  The subtracted distribution (right) reveals a double ridge structure~\cite{doubleridge}.}\label{fig:doubleridge}
\end{figure}

\begin{figure}[t!]
\centering 
\sidecaption
\includegraphics[width=0.5\textwidth]{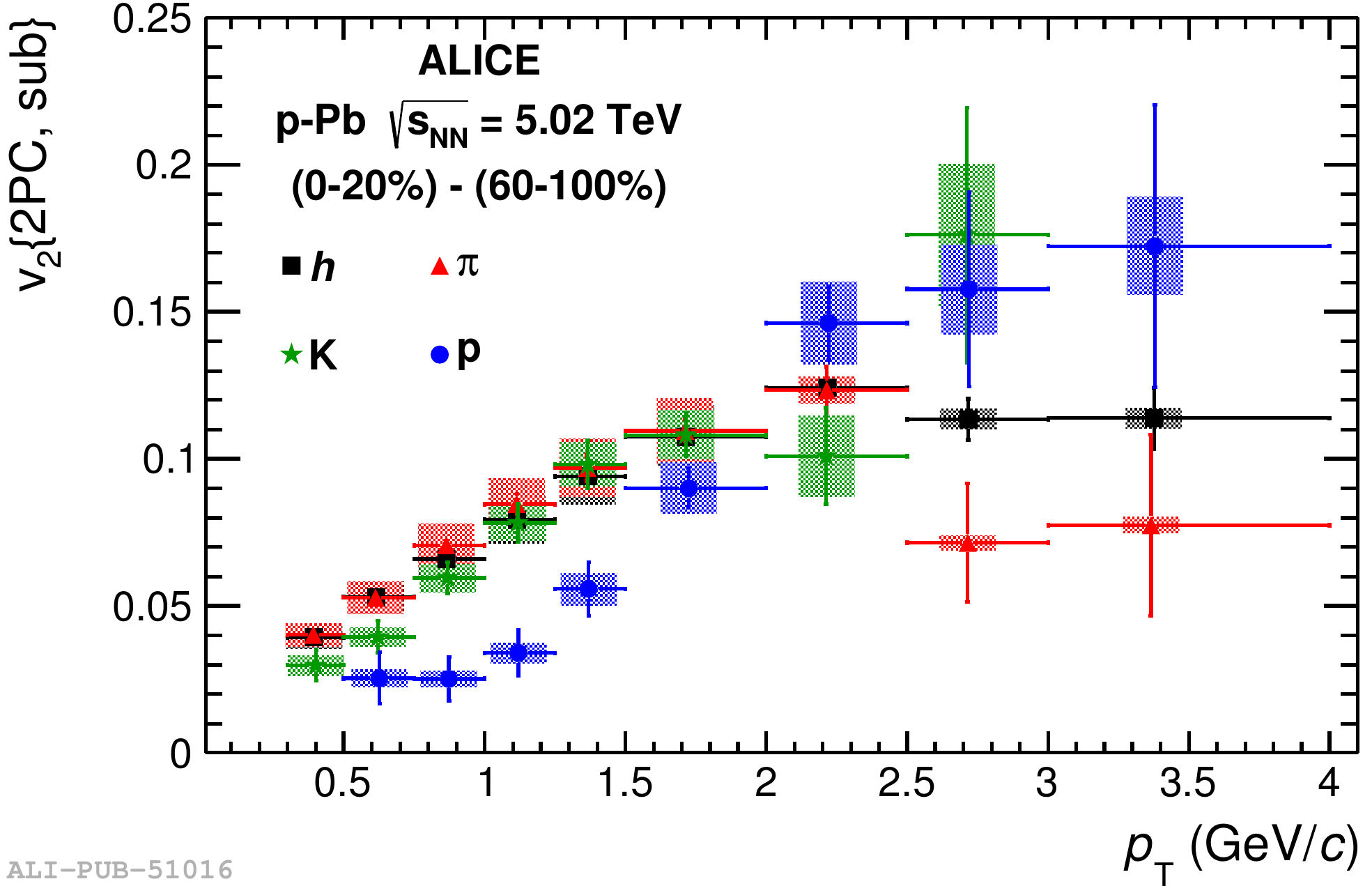}
\caption{The $v_2$ of $\pi$, $K$, and $p$ as well as inclusive unidentified hadrons was measured as a function of $\ptn{}$ in \pPb{} collisions at $\sqrt{s_{\mathrm{NN}}}$ = 5.02 TeV with the subtraction method~\cite{PIDridge}.\label{fig:fig4}}
\end{figure}

It was therefore surprising when a nearside ridge was observed in high multiplicity collisions of small systems, \pp{}~\cite{ppridge} and \pPb{}~\cite{pPbridge}.  Furthermore, it was observed that in \pPb{} collisions at $\sqrt{s_{\mathrm{NN}}}$ = 5.02 TeV the nearside peak yields are mostly independent of multiplicity~\cite{minijet}, meaning that for the same trigger and associated $\ptn{}$ the same jet population is selected regardless of multiplicity.  This served as justification to subtract the correlations in low-multiplicity events from the high-multiplicity correlation functions in order to remove correlations due to jet and minijet fragmentation.  This subtraction procedure (illustrated in Fig.~\ref{fig:doubleridge}) showed the nearside ridge more clearly and also revealed a symmetric ridge on the awayside~\cite{doubleridge,doubleridgeATLAS}.  This `double ridge' structure was decomposed into Fourier coefficients in order to extract the parameter $v_2$ in \pPb{} collisions.  The analysis was repeated with identified particles and it was observed that the $v_2$ shows similar mass ordering as was observed in \PbPb{} collisions.  Figure~\ref{fig:fig4} shows $v_2$ in \pPb{} collisions as a function of $\ptn{}$ for unidentified hadrons, pions, kaons, and protons~\cite{PIDridge}.  Results from CMS show similar behavior for $K^0_S$ mesons and $\Lambda$ baryons~\cite{PIDridgeCMS}.  The $v_2$ in \pPb{} collisions was also measured with the two- and multi-particle cumulant methods~\cite{cumulants,cumulantsATLAS,cumulantsCMS}.  It is important to note, however, that while the $v_2$ measured in \pPb{} collisions shows qualitatively similar features as $v_2$ measured in heavy ion collisions, the physical mechanism leading to a non-zero $v_2$ is still under theoretical debate and the presence of $v_2$ does \emph{not} necessarily imply the existence of hydrodynamics or a QGP in small collision systems.  

\subsection{Muon-hadron correlations}

In order to gain more information about potential collective effects and constrain theoretical calculations, it is important to measure the strength of the ridge to larger $\Delta\eta$ and to measure the dependence of $v_2$ on pseudorapidity.  Both of these points are addressed in the muon-hadron analysis performed in ALICE~\cite{muh}, in which correlation functions between muons at forward rapidities and charged hadrons at mid-rapidity are constructed in order to investigate the long-range behavior of the double ridge structure for $-5 < \Delta\eta < -1.5$.  

\begin{figure}
\centering
  \includegraphics[width=.36\linewidth]{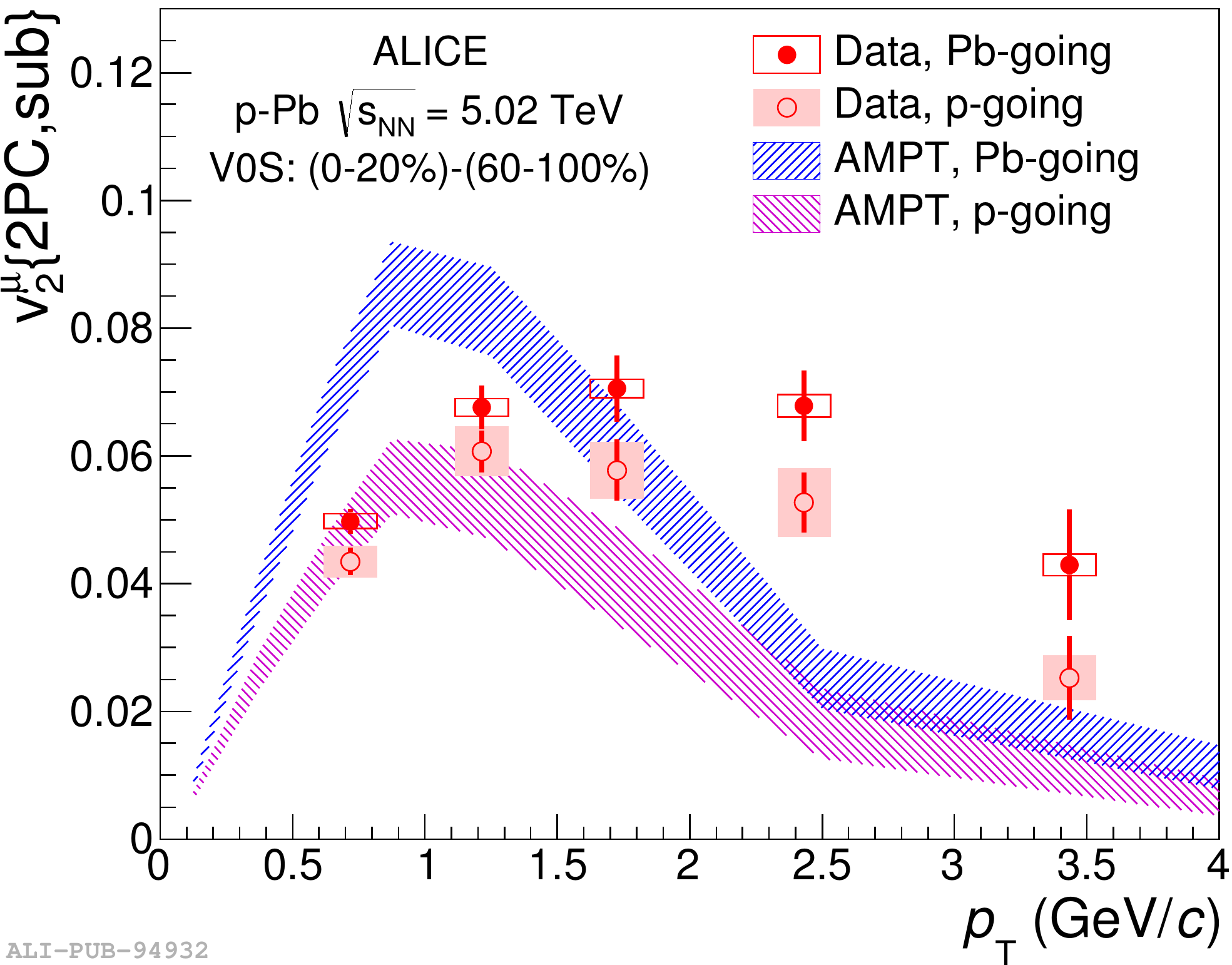}~
  \includegraphics[width=.36\linewidth]{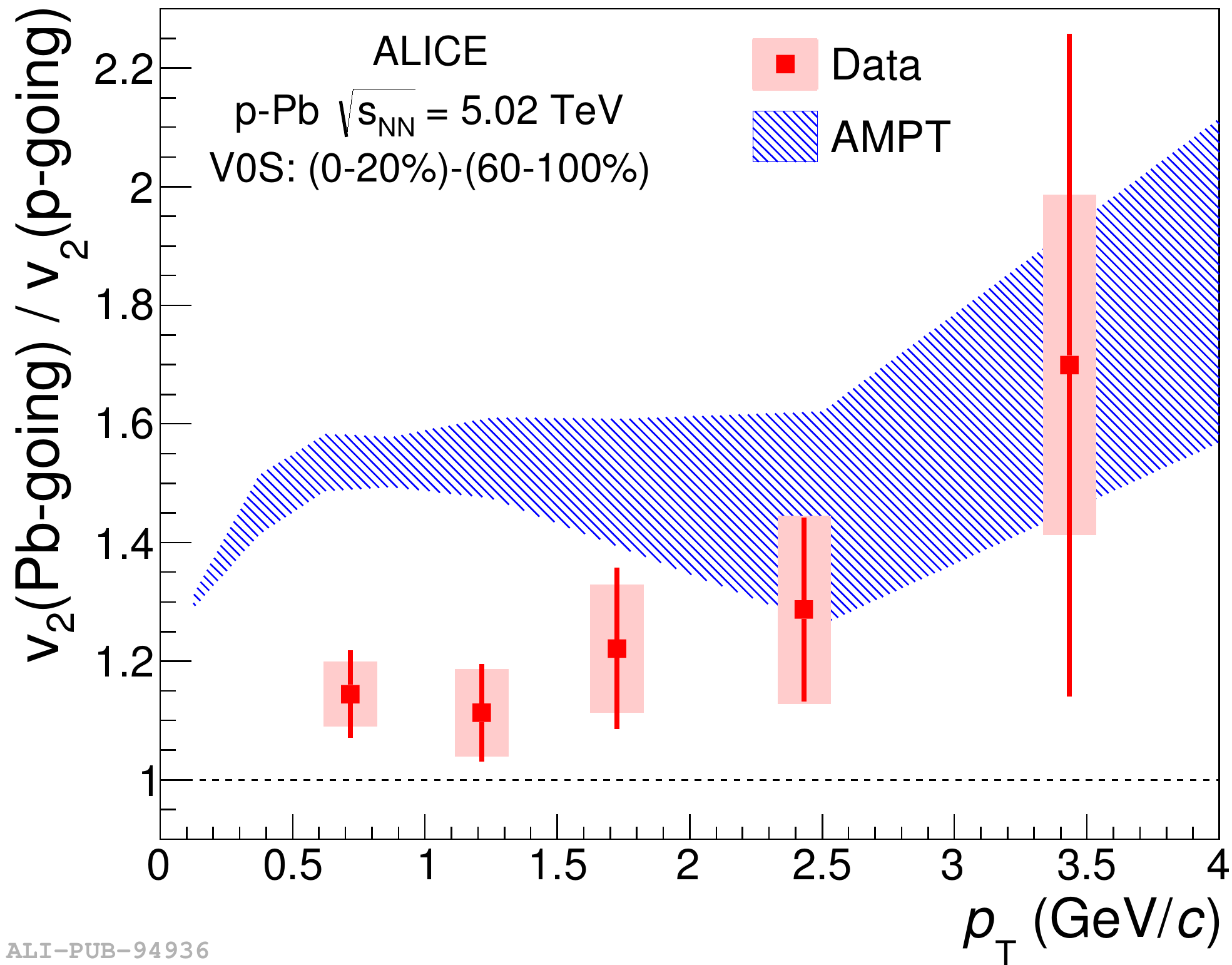}
  \caption{(left) The $v_2^{\mu}\{\mbox{2PC,sub}\}$ measured in the muon-hadron correlation analysis is shown in the p- and Pb-going directions and compared with results from AMPT.  (right) The ratio of the $v_2^{\mu}\{\mbox{2PC,sub}\}$ results in the Pb- and p-going directions is compared with AMPT.}
    \label{fig:v2ratio}
\end{figure}

The correlations between muons detected in the FMS and tracklets reconstructed in the ITS were measured in high-multiplicity (the top 20\% of the analyzed event sample) and in low-multiplicity (60-100\%) events.  As in~\cite{doubleridge}, the low-multiplicity correlations are subtracted from the high-multiplicity correlation functions to remove structures associated with jet fragmentation.  After subtraction, the correlation functions were projected onto $\Delta\varphi$, and then fit with a Fourier cosine series to extract $v_2$ for the muons detected at forward rapidities.  The resulting $v_2^{\mu}\{\mbox{2PC,sub}\}$ values are shown in Fig.~\ref{fig:v2ratio} for muons heading in the proton- and Pb-going directions.  The data are compared with an AMPT~\cite{refAMPT} simulation in which the muon decay products are scaled to account for the efficiency of the absorber in the ALICE FMS.  In Fig.~\ref{fig:v2ratio} it is seen that while AMPT qualitatively describes the $\ptn$-dependence at low $\ptn$, there are significant quantitative differences in the $\ptn$-dependence and $\eta$-dependence between data and the model.  At high $\ptn$ (above $\ptn \sim 2~\mbox{GeV}/c$), where muon production is dominated by heavy flavor decays, AMPT does not describe the data well.  This could be because heavy flavor muons have a non-zero $v_2$, or the parent particle composition or $v_2$ values in data and AMPT are different.  The ratio of $v_2^{\mu}\{\mbox{2PC,sub}\}$ in the Pb-going and p-going directions is also shown in Fig.~\ref{fig:v2ratio} where it is observed to be independent of $\ptn$ within the statistical and systematic uncertainties.  A constant fit to the data points shows that the $v_2$ is $(16\pm 6)\%$ higher in the Pb-going than in the p-going direction.  These results are qualitatively in agreement with model predictions.  However, current theoretical calculations cannot be directly compared with experimental results, because the effects of the absorber are included in the experimental data (unfolding such effects could not be done in a model-independent way).  Future model calculations should use the efficiencies provided in~\cite{muh} in order to compare directly to the experimental results.  

\section{Conclusions}
Single-particle inclusive and two-particle correlation measurements are used to characterize the \pp{} and \pPb{} collision systems.  The charged-particle multiplicity density has been measured across a range of energies including the top LHC energy of $\sqrt{s}$ = 13 TeV.  The pseudorapidity dependence of $\dNdeta$ has been compared with MC generators in order to further tune the models.  The multiplicity distributions were also compared with models and demonstrate KNO scaling up to $z\sim 4$.  In two-particle measurements, long-range correlations in pseudorapidity are observed up to $\eta\sim 4$ and $\Delta\eta\sim 5$.  The presence of these correlations is reminiscent of \PbPb{} collisions where the structures are frequently attributed to hydrodynamic flow, with similar mass ordering being observed in both small and large systems.  The $v_2$ of forward muons in the Pb-going direction is observed to be higher than in the p-going direction.  While these features are similar to correlations observed in heavy-ion collisions, further theoretical and phenomenological investigations are needed before any inferences about collectivity in small systems can be drawn.

%
\bibliography{biblio}

\end{document}